\documentclass{camera}
\usepackage{epsfig,a4}
\newcommand{\be}{\begin{equation}}
\newcommand{\ee}{\end{equation}}

\def\ga{\mathrel{\raise.3ex\hbox{$>$\kern-.75em\lower1ex\hbox{$\sim$}}}}
\def\la{\mathrel{\raise.3ex\hbox{$<$\kern-.75em\lower1ex\hbox{$\sim$}}}}

\newcommand{\tanb}{\ensuremath{\tan\beta}}

\newcommand{\PChiz}[1]{\ensuremath{\chi^0_{#1}}}
\newcommand{\lsp}{\ensuremath{\chi^0_{1}}}
\newcommand{\PSnu}{\ensuremath{\widetilde{\nu}}}

\newcommand{\Mlsp}{\ensuremath{m_{\PChiz{1}}}}

\newcommand{\mhalf}{{\ensuremath {m_{1/2}}}}

\begin{document}

\begin{picture}(160,1)
\put(310, 15){\parbox[t]{45mm}{{\tt ROM2F/2001/19}}}
\put(310,  2){\parbox[t]{45mm}{{\tt hep-ex/0106015}}}
\put(-40,-620){ 
{\it {\small Talk presented at the XIII$^{\mathrm{th}}$ 
Italian meeting on Physics at LEP, LEPTRE, Rome, April 2001.}}}

\end{picture}

%
\begin{center}
{\LARGE
SUPERSYMMETRIC DARK MATTER IN THE LIGHT OF LEP

\vspace{.5cm}

%

{\large 
G. Ganis

INFN - Roma II

}
}

\end{center}

%
%
\begin{abstract}
The negative outcome of searches for supersymmetry performed at LEP
have been used to derive indirect constraints on the parameters of 
the most plausible models for cold dark matter based on 
supersymmetric extensions of the Standard Model. 
The main results are summarized.
\end{abstract}

{\bf Introduction. }
The sensitivity of accelerators searches for new particles to 
possible solutions for the {\it Cold Dark Matter} (CDM) problem 
has been pointed out 
in Ref.~\cite{EllisLSP}a.
In the following we will enumerate the achievements which can be 
ascribed to the interpretation of the LEP results.
A discussion of the experimental problematics can be found in 
Ref.~\cite{Ganis2000}. We assume the reader being familiar with 
the naming scheme and the basic concepts intervening in supersymmetric 
models of elementary particles. 

{\bf Supersymmetric CDM candidates.}
We consider two models: a generic Minimal Supersymmetric extensions 
of the Standard Model (MSSM, see \cite{Nil84Hab85Mar97GDR99}), with R-parity 
conservation and unification of gaugino and scalar masses at unification 
scale, and a more constrained version of it (CMSSM or mSUGRA) in 
which the electroweak vacuum is required to be consistent with the 
unification relation, and the Higgs boson masses and the scalar trilinear 
couplings to unify at grand scale. 
There are two possible situations in which the Lightest Supersymmetric 
Particle (LSP) is a interesting CDM candidate 
\footnote{Additional CDM contributions could come also from the 
gravitino; however, the impact of the LEP results on gravitino
cosmology is marginal.}:
{\it i)} the LSP is  
a {\it sneutrino}, $\PSnu$, with mass in the ranges  
$\mathrm{M_{\PSnu}}\sim$few GeV/c$^2$ or $\mathrm{M_{\PSnu}}$ in 
[550,2300] GeV/c$^2$ (\cite{Ellis1998}); the latter 
range is excluded by direct searches (\cite{Caldwell1988}); 
{\it ii)} the LSP is the {\it lightest neutralino}, $\lsp$   
(a flexible candidate: see, for instance, 
Ref.~\cite{EllisLSP}b and references therein).

{\bf The first LEP result.}
The agreement with the predictions of the most recent determination 
of the Z widths (\cite{LEWWG2000}) sets 95\% confidence level upper 
limits of 6.2 and 1.7 MeV on new contributions to the total and invisible 
Z widths, respectively. 
Consequently, {\it sneutrinos masses up to about 40 GeV/c$^2$ are 
excluded definitely 
ruling out the sneutrino as supersymmetric candidate for CDM};
this is a good example of complementarity between indirect and 
direct searches. 

\begin{figure}[h]
 \begin{center}
  \begin{tabular}{c}
  \mbox{\epsfig{file=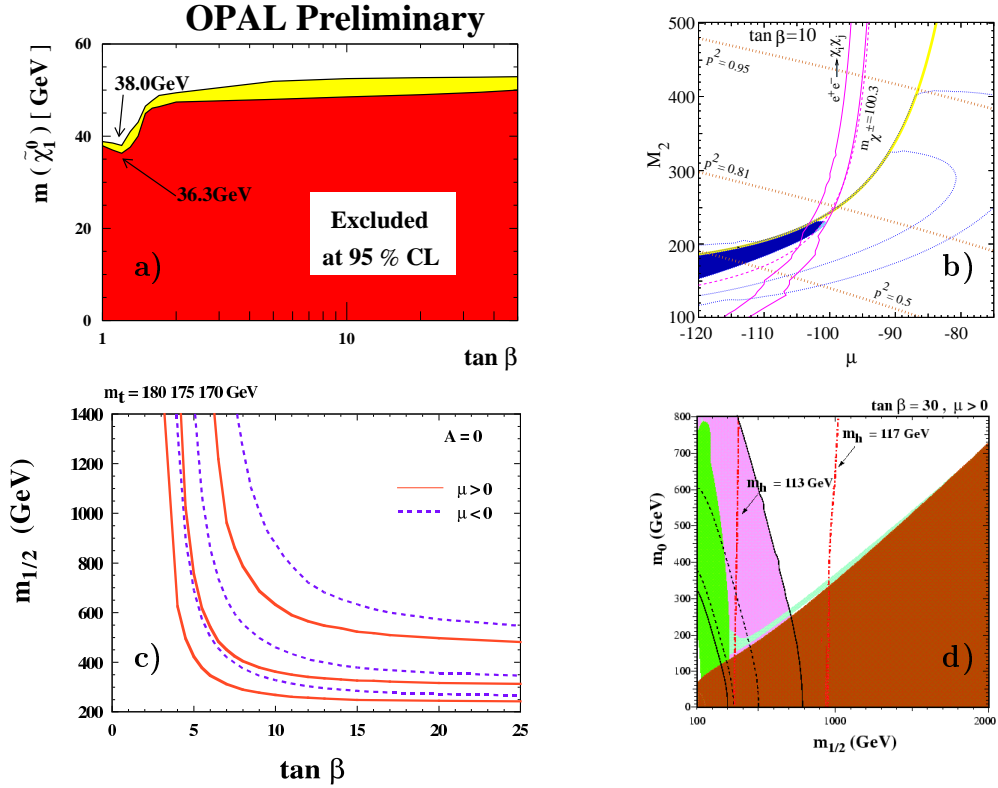,width=11.cm}}
  \end{tabular}
  \caption{\em {\small {\bf a)} Absolute limit on the $\lsp$ mass obtained 
by the OPAL collaboration in the MSSM using all the data 
sample; from Ref.~\cite{OPAL2001}. 
{\bf b)} Allowed regions (dark gray domains) 
in the {\it higgsino} corner of the 
$\mu,M_2$ plane for $\tanb=$10; from Ref.~\cite{EllisLSP}b. 
{\bf c)} Lower bounds on $\mhalf$ from the Higgs search in the 
CMSSM for different model parameters; from Ref.~\cite{EllisLSP}c. 
{\bf d)} Consistency of LEP Higgs ``signal'' 
(vertical dash-dotted lines) with new measurement of 
(g-2)$_\mu$ 
(medium-light gray, dashed, full lines)
in the CMSSM for $\tanb$=30 and $\mu\!>\!0$; 
from Ref.~\cite{EllisLSP}d.  
}}
  \label{figure}
 \end{center}
\end{figure}

{\bf The second LEP result.} 
The interplay of the most recent results of LEP searches for 
supersymmetric particles (charginos, neutralinos, sleptons, 
squarks) is discussed in Ref.~\cite{Azzurri2001}. 
The absence of any convincing 
evidence for a signal so far has allowed only to derive 
constraints on the parameter space. 
Interpreted as an absolute lower limit on $\Mlsp$, these look like 
in Figure~\ref{figure}a (from Ref.~\cite{OPAL2001}): in the MSSM, 
{\it neutralino masses smaller than $\sim$40 GeV/c$^2$ are 
disfavoured.} 

{\bf The third LEP result.} 
Figure~\ref{figure}b shows the exclusion bounds 
as a function of the {\it higgsino} content 
\footnote{Here $p\!=\!\sqrt{1\!-\!c^2_\gamma-c^2_z}$ with  
$c_\gamma$ ($c_z$) the photino (zino) component.}
$p$ of the $\lsp$: 
in the MSSM {\it the neutralino LSP cannot be predominantly higgsino}
(see Ref.~\cite{EllisLSP}b). 
This is in agreement with the predictions of the CMSSM. 

{\bf The forth LEP result.} 
The large radiative corrections to Higgs masses establish, in the 
CMSSM, a strong  relation between the Higgs sector and the gaugino 
masses. Taking into account the cosmology constraint 
\footnote{The requirement that the $\lsp$ relic density is in the 
right range.}, 
the lower limit on the Higgs boson mass can be translated 
into a lower limit on the common gaugino mass $\mhalf$ as shown in Figure 
\ref{figure}c. In the CMSSM, 
{\it tan$\beta$ value smaller than $\sim 5$ and neutralino masses 
smaller than $\sim$85 GeV/c$^2$ are disfavoured} (see Ref.~\cite{EllisLSP}c)
\footnote{
If the LEP ``signal''~\cite{LEPhiggs} is confirmed, an upper limit 
can also be set.}.

{\bf Conclusions.} 
The extensive searches performed at LEP for new phenomena have 
ruled out a large fraction of the MSSM parameter space interesting 
for CDM, basically limited 
only by the available centre-of-mass energy. 
In a CMSSM folded with the cosmology constraint,
the LEP results are compatible with the 
excess observed in the measurement of 
the muon anomalous magnetic moment (\cite{newmeas}) 
only for $\mu$ positive and for a small region of the parameter space 
(Figure~\ref{figure}d, from Ref.~\cite{EllisLSP}d). 
If confirmed, this excess may be seen as the first evidence 
for supersymmetry. 
A definite answer requires both a sizeable reduction of the 
uncertainty on hadronic 
contribution to the vacuum polarization (\cite{newmeas})
and {\it sparticle} detection at the LHC or at a TeV 
Linear Collider (\cite{EllisLSP}d).

%
\end{document}